\documentclass[a4paper,fleqn,usenatbib]{mnras}

\usepackage[T1]{fontenc}
\usepackage{ae,aecompl}

\usepackage{graphicx}	
\usepackage{amsmath}	
\usepackage{amssymb}	

\title[Tidbinbilla 70-m On-The-Fly mapping]{Implementation of Tidbinbilla 70-m On-The-Fly mapping and Hydrogen radio recombination line early results}

\author[G. F. Wong et al.]
{G.~F. Wong$^{1,2}$\thanks{E-mail: g.wong@westernsydney.edu.au},
 S.~Horiuchi$^3$,
 J.~A. Green$^{2,4}$, 
 N.~F.~H. Tothill$^1$,
 K. Sugimoto$^5$,\and 
 M.~D. Filipovic$^1$\\
$^{1}$Western Sydney University, Locked Bag 1797, Penrith 2751 NSW, Australia\\
$^{2}$CSIRO Astronomy \& Space Science, Australia Telescope National Facility, PO Box 76, Epping, NSW 2121, Australia\\
$^{3}$CSIRO Astronomy \& Space Science/NASA Canberra Deep Space Communication Complex, PO Box 1035, Tuggeranong ACT 2901, Australia\\
$^{4}$SKA Organisation, Jodrell Bank Observatory, Lower Withington, Macclesfield SK11 9DL, UK\\
$^{5}$National Astronomical Observatory of Japan, 2-21-1 Osawa, Mitaka, Tokyo 181-8588, Japan
}

\date{Accepted XXX. Received YYY; in original form ZZZ}

\pubyear{2015}

\begin{document}
\label{firstpage}
\pagerange{\pageref{firstpage}--\pageref{lastpage}}
\maketitle

\begin{abstract}
On-the-fly mapping of cm-wave spectral lines has been implemented at the 
the Tidbinbilla 70-m radio antenna.
We describe the implementation and data reduction procedure and present new H92$\alpha$ radio recombination line maps towards Orion\,A and Sagittarius\,A.
Comparison of the Orion~A map to previous observations suggests that the lines arise largely from gas with electron density of 100--200\,cm$^{-3}$.
On-the-fly mapping is  very efficient at generating large maps of bright lines (such as radio recombination lines), but will still yield strong efficiency gains for smaller maps of fainter lines, such as the ammonia inversion lines at the 1.3\,cm wavelength.
\end{abstract}




\section{Introduction}
\label{sec:intro}

The Deep Space Station 43 (DSS-43) is a 70-m single dish radio telescope that is part of the NASA Canberra Deep Space Communication Complex (CDSCC) located in the Australian Capital Territory (hereafter referred to as `Tid-70m').
A small fraction of its time is devoted to radio astronomy and it is important to maximise the scientific return on this time. 
As such developments have been under way to implement On-The-Fly (OTF) mapping capabilities, to allow for an efficient survey option.
Following the preliminary work of \cite{young}, we present full details of the implementation of OTF mapping at the telescope allowing for greater mapping efficiency.
More efficient spatial mapping of the ammonia inversion transitions in the 17--27\,GHz K-band is a strong driver of the implementation of OTF on the Tid-70m, but the technique may be used to map other radio lines, such as radio recombination lines (RRLs) that fall in
the cm-wave bands.
We used OTF mapping with Tid-70m to observe the H92$\alpha$ RRL at 8.3\,GHz towards the prominent Galactic {\sc Hii} region Orion\,A (Ori\,A) and the Galactic Centre feature Sagittarius\,A (Sgr\,A).

OTF mapping involves sampling the sky while the antenna is moving at a constant rate, rather than integrating at a discrete position on the sky, then moving to the next position. The technique is widely used by mm- and submm-wave telescopes to produce large-scale maps in molecular transitions and the greybody continuum, such as CO~3--2 mapping towards M\,83 \citep{aste_co}, maps of CO and 1.1\,mm continuum emission towards Ori\,A  \citep{cont}, and 90\,GHz maps of star-forming clumps \citep{jackson}. Detailed descriptions of the technique have been presented by \citet{mang} and \citet{saw}. 
Implementation of OTF mapping at the Australia Telescope National Facility (ATNF) Mopra telescope has yielded several large mapping surveys \citep[e.g.][]{magma, jones, mopragal}.

\cite{2012MNRAS.425.2352P} reviewed the physics of hydrogen RRL emission: the populations depart significantly from local thermodynamic equilibrium (LTE) over frequency ranges that depend on the electron density. \citeauthor{2012MNRAS.425.2352P} consider ALMA and EVLA observations of compact high-density \textsc{Hii} regions, but also show that for the larger, less dense \textsc{Hii} regions that might be mapped with a single-dish telescope, non-LTE effects are important in the centimetre-wavelength range. \cite{RRL_book} used a non-LTE analysis to show how the line brightness varies with frequency and electron density. We may therefore estimate the electron density from the frequency dependence of line brightness.
The line brightness depends on electron temperature and density, so RRL maps can be used to infer the structure of ionised gas \citep[e.g.][]{1978ApJ...226..869J}.
\section{Observing process}
\label{obs_pro}

\begin{figure}
\begin{center}
\includegraphics[width=0.46\textwidth]{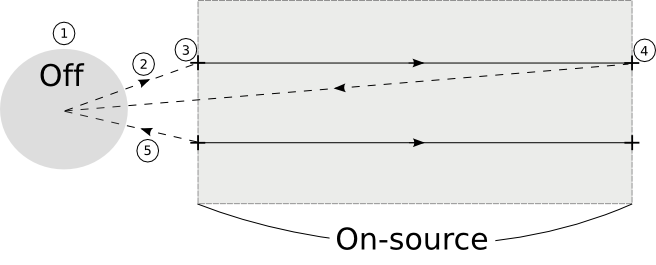}
\caption{
The sequence of steps in an OTF observation.
The target region to be mapped is shown as a shaded rectangle.  
The solid line indicates telescope movement, starting and stopping at the crosshairs, during which the spectrometer is continuously sampling. 
The dashed line shows telescope movement without spectrometer sampling. 
In this example, every scan row is observed before making an OFF measurement, shown as a shaded circle.
}
\label{pattern}
\end{center}
\end{figure}

While many OTF scan patterns can be implemented, we have used the simplest, the raster (Fig.~\ref{pattern}), which is optimised for mapping rectangular regions.
Raster scanning moves the telescope on-source, `ON', in a straight line, then offsets by a fraction of a beam, perpendicular to the scan direction, before moving the telescope in the opposite direction.
An emission-free reference position, `OFF', is also observed to correct for the bandpass. 
As the telescope moves across the source, the spectrometer samples data at regular intervals.

Our implementation is a variation of the raster geometry, where the order of observing on-source or reference position is flexible: an OFF can be observed before or after an ON, or after multiple ON scans (see \S\ref{sec:ant} and \ref{sec:spec} for details).
Maps can be made in either ($\alpha\cdot\cos\,\delta_{0}$, $\delta$) or (\textit{a}$\cdot\cos e_0$, $e$) coordinate systems, with the scan direction parallel to either coordinate axis \footnote{$a, e$ are azimuth and elevation. $\delta_{0}$ is the declination of the starting position of the observation. The longitudinal coordinates are referred to as cross-Dec and cross-El in the telescope control system.}.


\subsection{Software Architecture}
\label{sec:ctrl_sw}
\begin{figure}
\begin{center}
\includegraphics[width=0.5\textwidth]{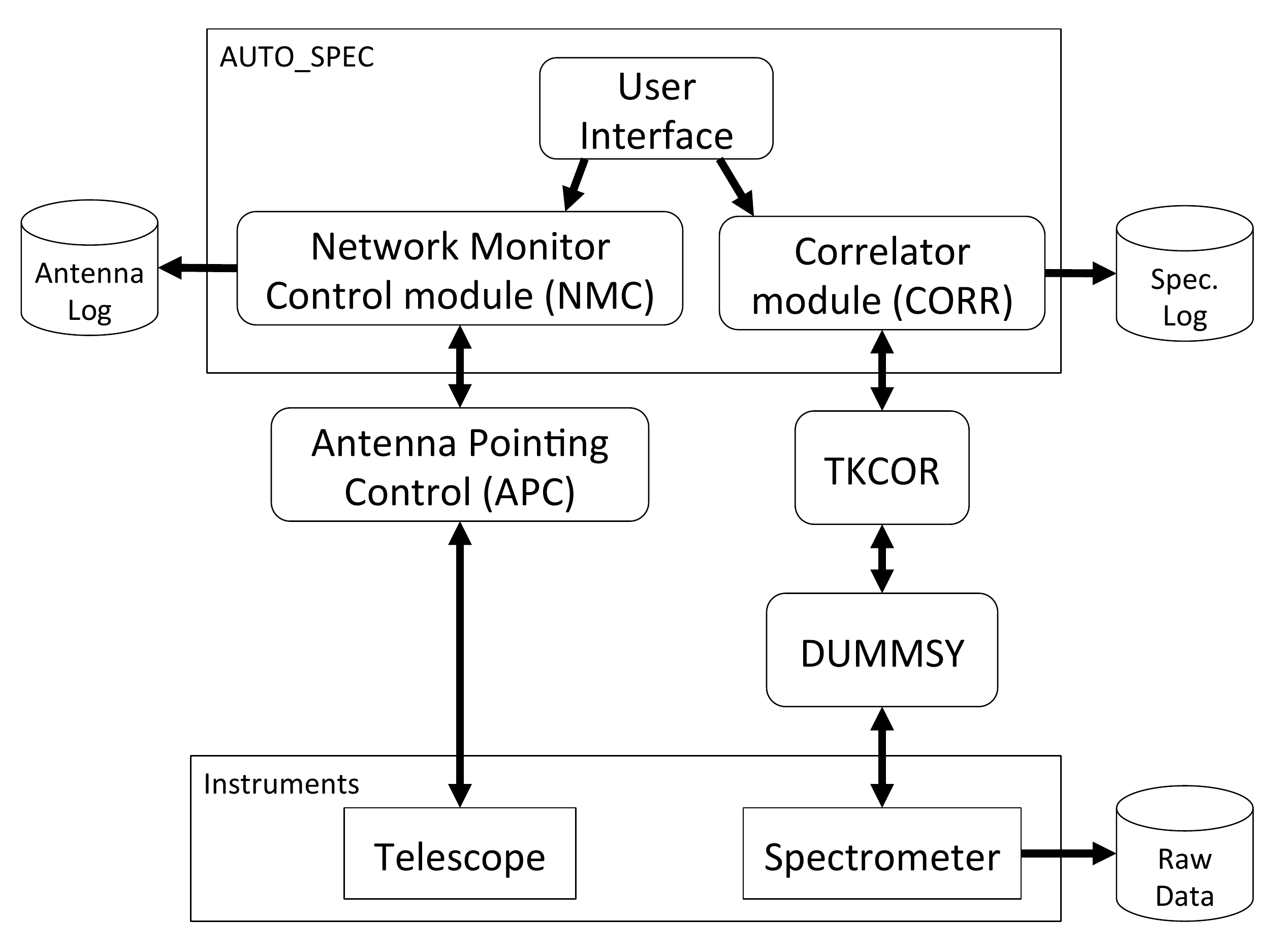}
\caption{Schematic illustration of the telescope system. 
The rounded rectangles are processes involved with OTF, rectangles are individual instruments and cylinders represent data storage and files.}
\label{schematic}
\end{center}
\end{figure}

The OTF observing mode was implemented on the control software \textsc{auto\_spec}, a \textsc{perl} based script that acts as an interface to the telescope and spectrometer.
This software sends commands via two \textsc{perl} modules: network monitor control (\textsc{nmc.pm}) and correlator (\textsc{corr.pm}). 
These commands are sent to the antenna pointing control (\textsc{apc}) and the user interface (\textsc{tkcor}) respectively.
The \textsc{nmc} module interacts with NASA controlled modules of the telescope.
The commands the \textsc{nmc} module sends are: setting position; offset; slew rate; stop and stow. 
The module also retrieves information on the current {\it a} and {\it e} coordinates, offsets, slew rate and supplementary information (temperature, pressure, humidity, wind speed, wind direction, precipitation and time).
The antenna logs that are generated by the \textsc{nmc} module contain telescope commands and records of the position.
The \textsc{corr} module interacts with the user interface \textsc{tkcor}, a \textsc{Perl/Tk} module that is the interface of the process \textsc{dummsy}.
\textsc{dummsy} sends commands to the physical correlator.
Logs from the spectrometer are generated from \textsc{corr.pm}.
Fig.~\ref{schematic}. is a schematic illustration of the architecture of the software control system.

\subsection{Antenna}
\label{sec:ant}

The commands to move the telescope come from the schedule file, setting the parameters of the observation, containing positional information for the reference position and starting position on source, time on source and frequency.
Scanning axis and direction for on-source scanning are programmed into the control software \textsc{auto\_spec} before observations begin.
In a typical OTF mapping observation, where a reference position is taken, first, then a single on source observation:

\begin{enumerate}
	\item OTF mapping begins with integrating on an emission-free reference position;
	\item the telescope slews to the start position of the target source;
	\item the OTF scan starts with \textsc{auto\_spec} commanding the telescope control to settle on the $\alpha$ and $\delta$ starting point before moving the telescope in a set direction at a constant rate with a timer starting;
	\item the scan rate of the telescope is calculated based on the beam size, and sampling rate:
\begin{equation} 
R = \frac{\theta_{int}}{t_{int}} = 
\frac{d\alpha}{dt}\cos\,\delta_0
\label{scanRate}
\end{equation}
(assuming scanning in $\alpha\cos\,\delta_{0}$), where $R$ is the scanning rate of the telescope in degrees\,s$^{-1}$, $\theta_{int}$ is defined as $\theta_{FWHM}/n$ ($n$ is the number of times to sample the beam, generally $\geq 3$) and ${t_{int}}$ is the number of seconds per integration;
	\item  as the telescope continues to slew across the source, the antenna log samples the telescope's position in $a$ and $e$ every 15 seconds.
	\item the telescope will be sent the command to stop when the timer matches the on source time in the schedule file;
	\item once the telescope has stopped, the next line in the schedule file is read.

\end{enumerate}

\subsection{Spectrometer}
\label{sec:spec}
The Tid-70m is equipped with a Parkes Multibeam correlator block \footnote{\url{www.atnf.csiro.au/observers/tidbinbilla}}.
OTF observations can use the full 64~MHz bandwidth with two polarisation products of up to 2048 channels for each polarisation.
As the telescope is being driven, the spectrometer is integrating the spectra over a minimum time interval of three seconds.
The data from the spectrometer are then written to an ATNF format file (\textsc{rpfits}\footnote{\url{www.atnf.csiro.au/computing/software/rpfits.defn}}) along with time stamps, positional information, frequencies and summary information of the observation.
The summary information contains start and finish time, observation project number, source name, starting position in $\alpha$ and $\delta$ as well as {\it a} and {\it e}, number of Intermediate Frequencies (IFs), observed frequency, reference frequency and system temperature ($T_{sys}$/K).
The spectrometer will only record the starting position of each row of the map into the spectrometer log file through the correlator module, and this starting position is derived from the schedule file, not the telescope encoders.
The average position of each sample was therefore derived by cross-checking the \textsc{nmc} log against the spectrometer log, taking into account slew times and correlator delays.
This was done by the python data processing scripts \textsc{fillobservatories.py}\footnote{\url{spacescience.uws.edu.au/~honours/Graeme_W/fillobservatories.py}} and \textsc{fixDirection\_v2.py}\footnote{\url{spacescience.uws.edu.au/~honours/Graeme_W/fixDirection_v2.py}}. 
These scripts also adjust the sample positions (see section~\ref{sec:reduction}).

\subsection{Data processing}
\label{sec:reduction}
Data processing of 70m-Tid OTF data uses the Common Astronomy Software Applications \citep[\textsc{casa} version 4.1 and 4.2;][]{casa}.
The raw file (\textsc{rpfits}) containing header information, tables and observation data is imported into \textsc{casa}, with positions converted into radians.
Correcting the position of each integration is based on converting the recorded velocity taken from \textsc{nmc} log into radians and compensating for the initial delay caused by the correlator.
The scan rate ($R$), is then used to correct the position of each integration along the $\alpha\cos\delta_0$ axis:
%
%
\begin{equation} 
\alpha_{i} = \alpha_{0} + \frac{R}{\cos\,\delta_{0}}\cdot(t_{i}-t_{0})
\label{new_pos}
\end{equation}
where $\alpha_{0}$ and $t_0$ are the starting R.A.\ and time, and $\alpha_{i}$ and $t_i$ are the R.A.\ and time of the current integration ($\delta_{0}$ is the corresponding declination for $\alpha_{0}$).
The new positions are written into the scan table, as well as the recalculated {\it a} and {\it e} for each new position.
The antenna temperature $T_{A}$ is obtained by bandpass correcting the raw spectrum:
\begin{equation} 
T_{A}  = T_{sys}\cdot \frac{ON-OFF}{OFF}
\label{cal_quotient}
\end{equation}
$T_{sys}$ is the system temperature, $ON$ is the on-source spectrum and $OFF$ is the reference spectrum.

The schedule file contained a reference scan for each on-source scan; the \textsc{sdcal} task in \textsc{casa} assumes this format.
Polynomial baseline subtraction can occur at this stage, or a continuum subtraction can be applied after the imaging stage is completed to preserve the continuum emission.
Smoothing functions such as the Hanning function can also be applied if the target spectral line is narrow.
Calibrated data are then exported into a measurement set (MS) format for the imaging stage, which creates a position-position-velocity cube using both circular polarisations.
The gridding kernel is the default 2-D top-hat function with a 1-pixel width.
Continuum subtraction from the image cube can be applied (if baseline subtraction was not applied to the spectra), through the specification of a line-free channel range, resulting in a continuum cube and emission line cube.
The data cube can then be exported to \textsc{fits}.
The detailed \textsc{casa} data reduction procedure is described in Appendix~\ref{app:casa}.

\section{Test Observations}
\label{sec:obs}
The H{\sc{ii}} region Ori\,A and Galactic centre source Sgr\,A were the targets for our test observations (Table~\ref{tab:regionInfo}).
Observations were conducted under Tid-70m project T206, as part of the OTF development programme.
Our test observations used the 8.2--8.6~GHz X-band receiver and recorded both circular polarisations with 64~MHz bandwidth, sampling at a quarter FWHM with the minimum three second intervals.
The beam has FWHM 1.8\arcmin, and individual scans (map rows) were spaced by 30\arcsec\ in Dec, while the telescope scanned along the $\alpha\cdot\cos\,\delta_{0}$ axis.
A schedule file parameter set the time taken for the telescope to complete each scan: 
These observations used 120s-long scans with 30s integrations on the reference position.
The Hanning function was not applied for the maps presented, as RRLs are broad spectral features, so Hanning is not required.
Opacity correction was not applied to our maps, but attenuation due to opacity would be within 5\% at the lowest elevation during the 8\,GHz observations.
There was no comparison to a flux calibrator source, but our observations agree within 10\% with those of \cite{Cesarsky}.

\begin{table*}
\caption{Table showing the different sources observed, the date observed, the amount of time taken to observe, map size in arcmin, central position of the map and reference position.}
\begin{center}
\begin{tabular}{lcccccc}
\hline
Region&Date &Time&Map Size&Map Size& Central map position& Reference position\\
&Observed& Duration& (R.A. $\times$ Dec)&(Pixels)& R.A. (J2000) Dec (J2000)&R.A. (J2000) Dec (J2000)\\
\hline%
  Ori\,A  &  2013-09-24 & 2.4 h & 18.9\arcmin$\times$ 17.5\arcmin\ & 42 $\times$ 36 & 05:35:19.2 --05:18:47.7 & 05:40:30.0 --05:16:16.0\\
  Sgr\,A  &  2013-09-23 & 2.8 h & 20.4\arcmin$\times$ 16.0\arcmin\ & 42 $\times$ 42 & 17:45:58.9 --28:58:22.1 & 17:55:00.0 --28:00:00.0\\ 
 &  2013-09-24 & 1.3 h & 20.4\arcmin$\times$ 10.5\arcmin\ & 42 $\times$ 21 & 17:45:58.9 --28:43:28.3 & 17:55:00.0 --28:00:00.0\\ 
\hline
\end{tabular}
\end{center}
\label{tab:regionInfo}
\end{table*}

Ori\,A and Sgr\,A were mapped with the same telescope scan speed of $-2.48$\,mdeg\,s$^{-1}$ along $\alpha\cdot\cos\,\delta_{0}$; converting the scan speed to $\alpha$ and using the minimum integration time of 3s, Ori\,A and Sgr\,A have cell sizes of 
0.45\arcmin\  and 0.51\arcmin\ respectively along the scan direction, and 0.5\arcmin\ across the scan direction.
Offsets caused by the delay of the spectrometer were between 0.30\arcmin\ and 0.60\arcmin\ for Ori\,A and 0.34\arcmin\ and 0.68\arcmin\ for Sgr\,A.
The scan speed is taken from the \textsc{nmc} log which records telescope speed, and the integration time is taken from correlator logs (The negative scanning speed indicates that the telescope was moving in decreasing $\alpha\cdot\cos\,\delta_{0}$.). The effect of these corrections is to shift each scan row by a slightly different amount along the scan direction. The resulting Sgr\,A image has a continuum peak shifted to the position of Sgr\,A$^{*}$ compared to the uncorrected image which had a 2.5\arcmin\ offset from Sgr\,A$^{*}$.

\begin{figure}
\begin{center}
\includegraphics[trim = 0cm 0cm 0cm 0cm ,clip=true, width=0.5\textwidth]{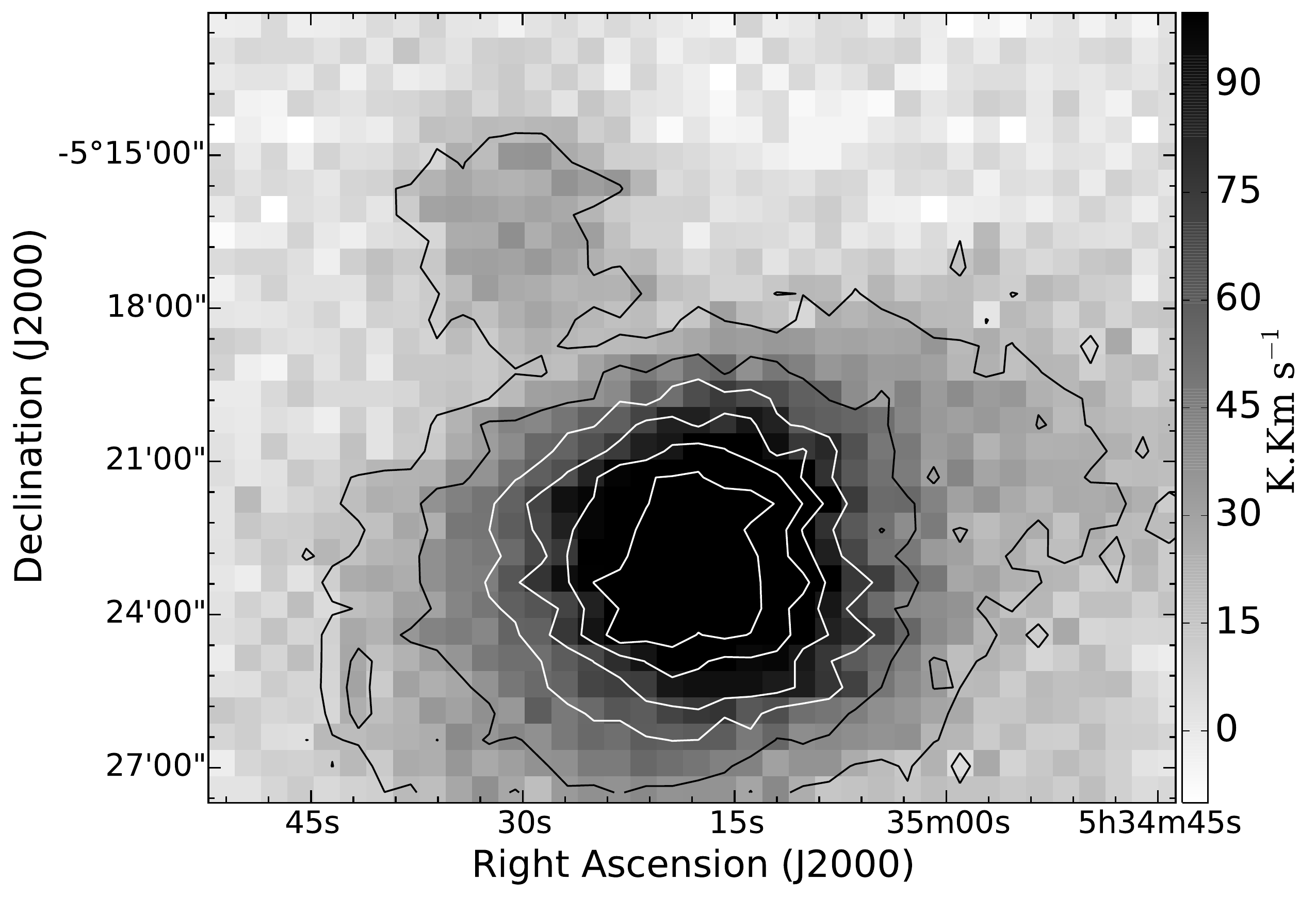}
\caption{
Integrated H92$\alpha$ emission map towards Ori\,A; contours at 20, 40, 60\dots 140~K\,km\,s$ ^{1} $. }
\label{fig:orionA}
\end{center}
\end{figure}

\begin{figure}
\begin{center}
\includegraphics[trim = 0cm 0cm 0cm 0cm ,clip=true, width=0.5\textwidth]{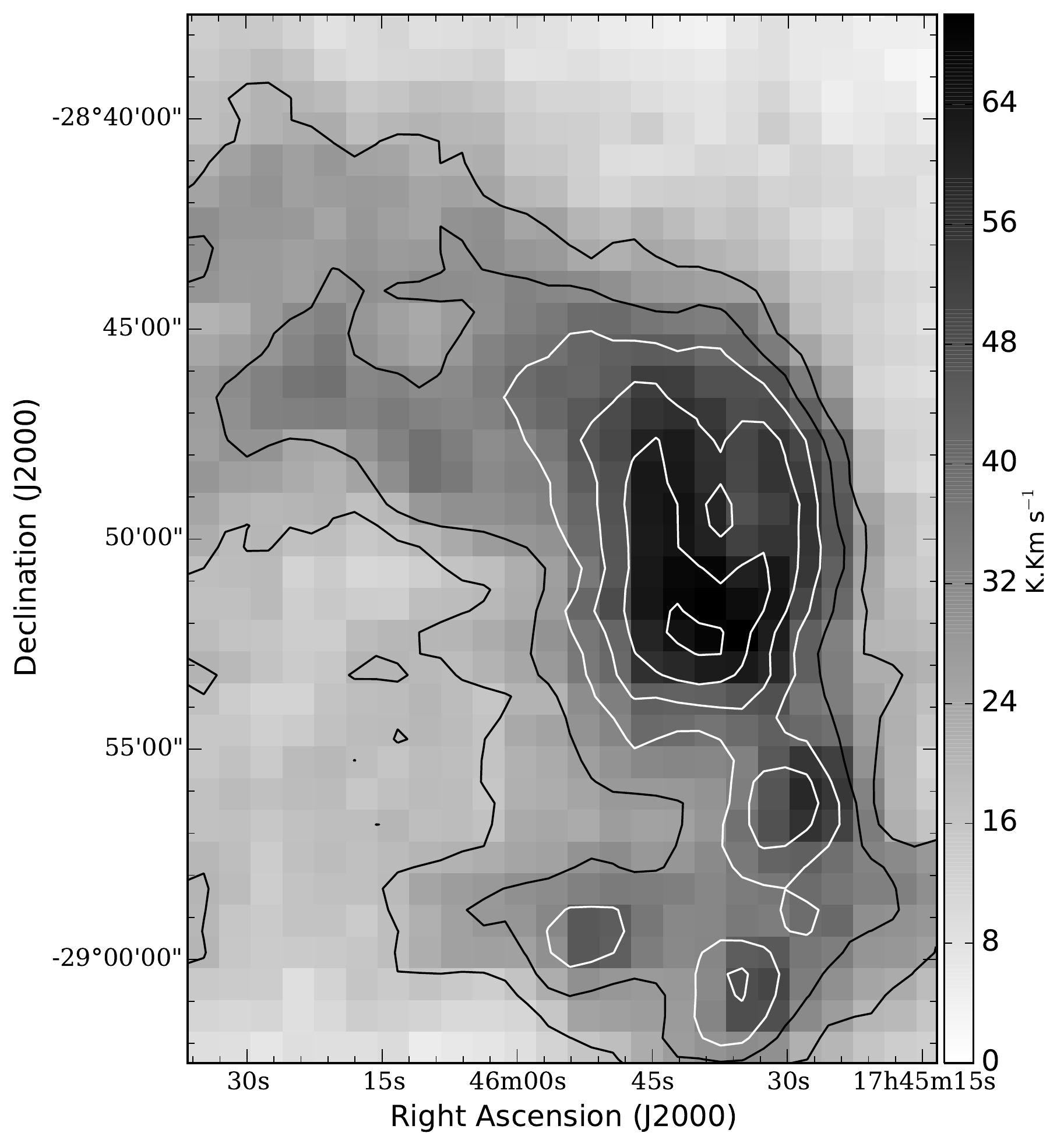}
\caption{
Integrated H92$\alpha$ emission map with incomplete continuum subtraction towards Sgr\,A; contours at 20, 30, 40\dots 70~K\,km\,s$^{-1}$.}
\label{fig:SGA}
\end{center}
\end{figure}

\begin{figure*}
\begin{center}
\includegraphics[width=1\textwidth]{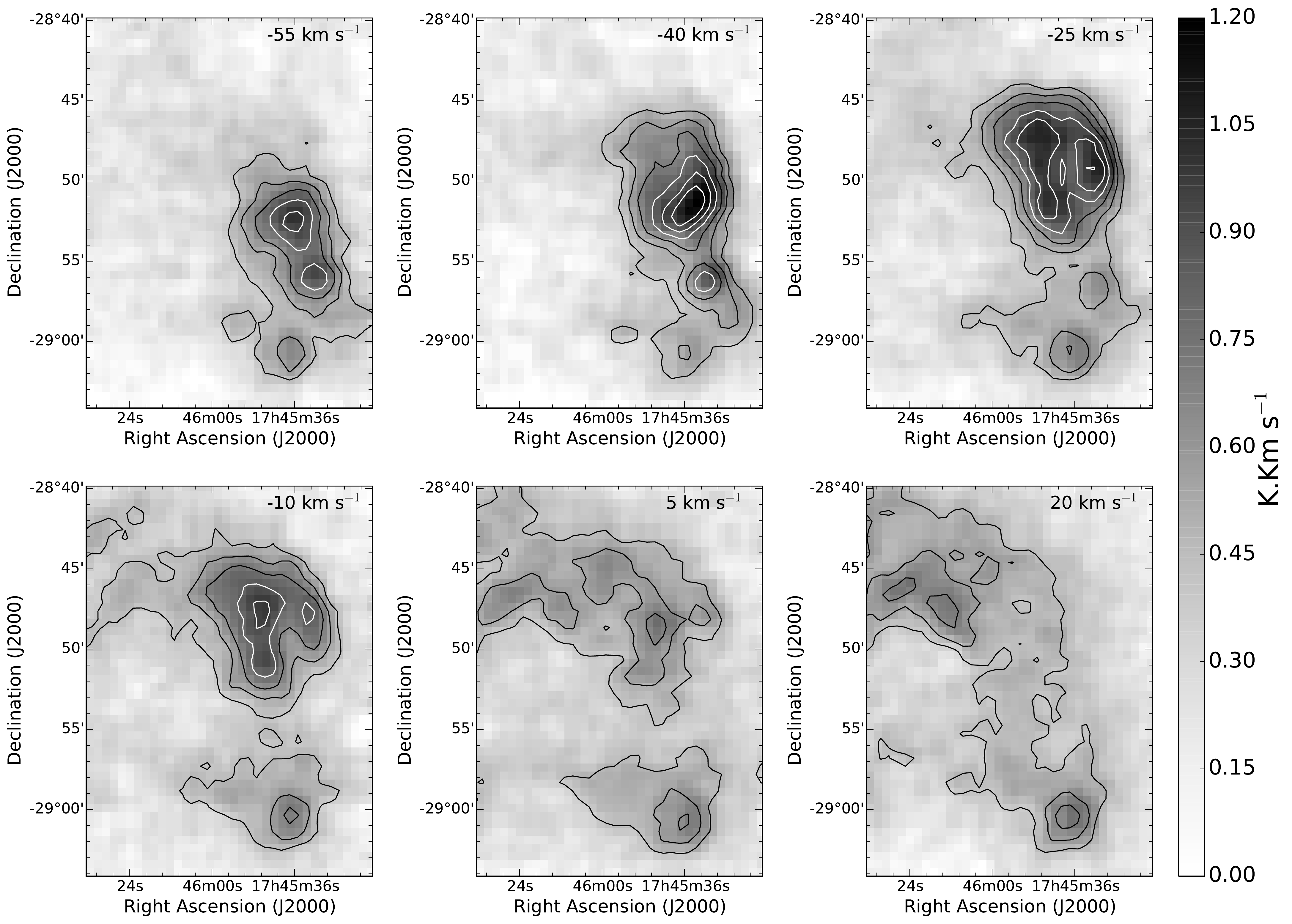}
\caption{On-the-fly channel maps with incomplete continuum subtraction of H92$\alpha$ emission towards Sgr\,A; 
contours at 0.3, 0.45, 0.6\dots 1.2\,K.} 
\label{fig:SGA_channel}
\end{center}
\end{figure*}

Both sources were observed with channels of width 1.13~km\,s$^{-1}$.
The Ori\,A integrated H92$\alpha$ emission map is shown in Fig.\,\ref{fig:orionA}.
Continuum subtraction was applied to the Ori\,A cube after the imaging process, as this correctly produced a flat spectrum.
Sgr\,A was observed on two separate days; data from each day were processed separately before merging the datasets during the imaging stage.
H92$\alpha$ integrated and channel maps towards Sgr\,A are shown in Figs.\,\ref{fig:SGA} and \ref{fig:SGA_channel} respectively.
While the data cube has a channel width of 1.13~km s$^{-1}$, the channel map in Fig.~\ref{fig:SGA_channel} uses a channel width of 15~km s$^{-1}$.
Because of the complex emission structure of the region, an attempt was made to select a line-free channel range to create a polynomial to remove continuum emission from the spectra.
Examining the attempted continuum subtraction, we see the Northern region with continuum emission successfully removed. 
However, around the Southern peak ($\delta<-28^{\circ}55\arcmin$) broadband emission can still be seen.
Continuum emission is still present within Figs.\,\ref{fig:SGA} and \ref{fig:SGA_channel}.

\section{Discussion}
\label{sec:dis}

\begin{table*}
\caption{Gaussian fits to sample H92$\alpha$ spectra.}
\begin{center}
\begin{tabular}{lcccccccc}
\hline
Region&R.A. &Dec& V$_{LSR}$ &$\Delta \nu$ & T$_{A}$ \\
&J2000& J2000& (km\,s$^{-1}$) & (km\,s$^{-1}$) & (K)\\
\hline%
  Ori\,A  &  05:35:17.8& --05:22:15.0&--2.7$\pm$0.3&25.0$\pm$0.7&5.07$\pm$0.12\\
  Sgr\,A  &  17:45:41.4& --28:52:14.9& --40.5$\pm$0.8& 56.2$\pm$2.0& 0.88$\pm$0.03\\
\hline
\end{tabular}
\end{center}
\label{tab:guass_para}
\end{table*}

\subsection{Orion A}

\begin{figure}
\begin{center}
\includegraphics[trim = 0cm 0cm 0cm 0cm ,clip=true, width=0.5\textwidth]{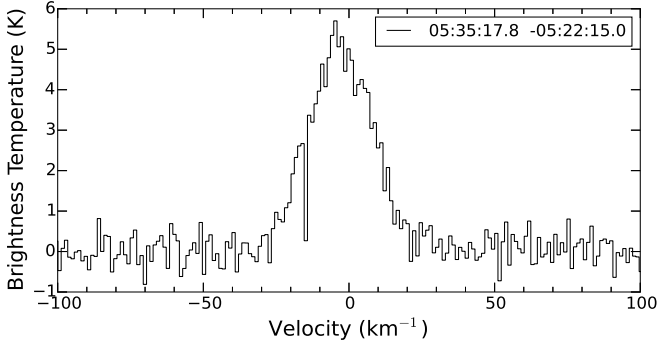}
\caption{H92$\alpha$ spectrum towards Ori\,A at the position of peak $T_L$ (coordinates top right).} 
\label{fig:OrionA_spectra}
\end{center}
\end{figure}

Our data constitute the first fully-sampled single-dish RRL map of Ori\,A, though sparsely-sampled maps made up of selected beam positions have previously been published by \citet{1978ApJ...226..869J} and by \cite{1988PASJ...40..581};
the RRL structure (Fig.~\ref{fig:orionA}) is similar to that seen in a 23\,GHz continuum map  \cite[resolution 42\arcsec,][]{1984A+A___138__225W}.
Many hydrogen RRL transitions have been observed with single-dish telescopes towards Ori\,A, mostly towards similar positions, so that the beams overlap.
Table~\ref{tab:lit_ref} lists H(\textit{n})$\alpha$ RRL transitions observed towards positions close to the peak brightness that we observe in Ori\,A.
The spectrum corresponding to the peak brightness in our H92$\alpha$ map (Fig.~\ref{fig:OrionA_spectra} and Table~\ref{tab:guass_para}) may be compared to a previous observation of the same transition by \citet{Cesarsky}, which used the
DSN Goldstone antenna, then 64\,m in aperture, with beamwidth of 2.5$\arcmin$.
No pointing coordinates are given, but the published line parameters
(T$_{A}$ = 4.65~K, V$_{LSR}$ = --2.7$\pm$2~km\,s{$^{-1}$}, $\Delta \nu$ = 26$\pm$2~km\,s{$^{-1}$}), agree with our results (Table~\ref{tab:guass_para}) within the error intervals,  except for line amplitude; but even here, the discrepancy of 0.42~K is less than 10\%.

\begin{figure}
\begin{center}
\includegraphics[trim = 1cm 0cm 1.5cm 1cm ,clip=true, width=0.5\textwidth]{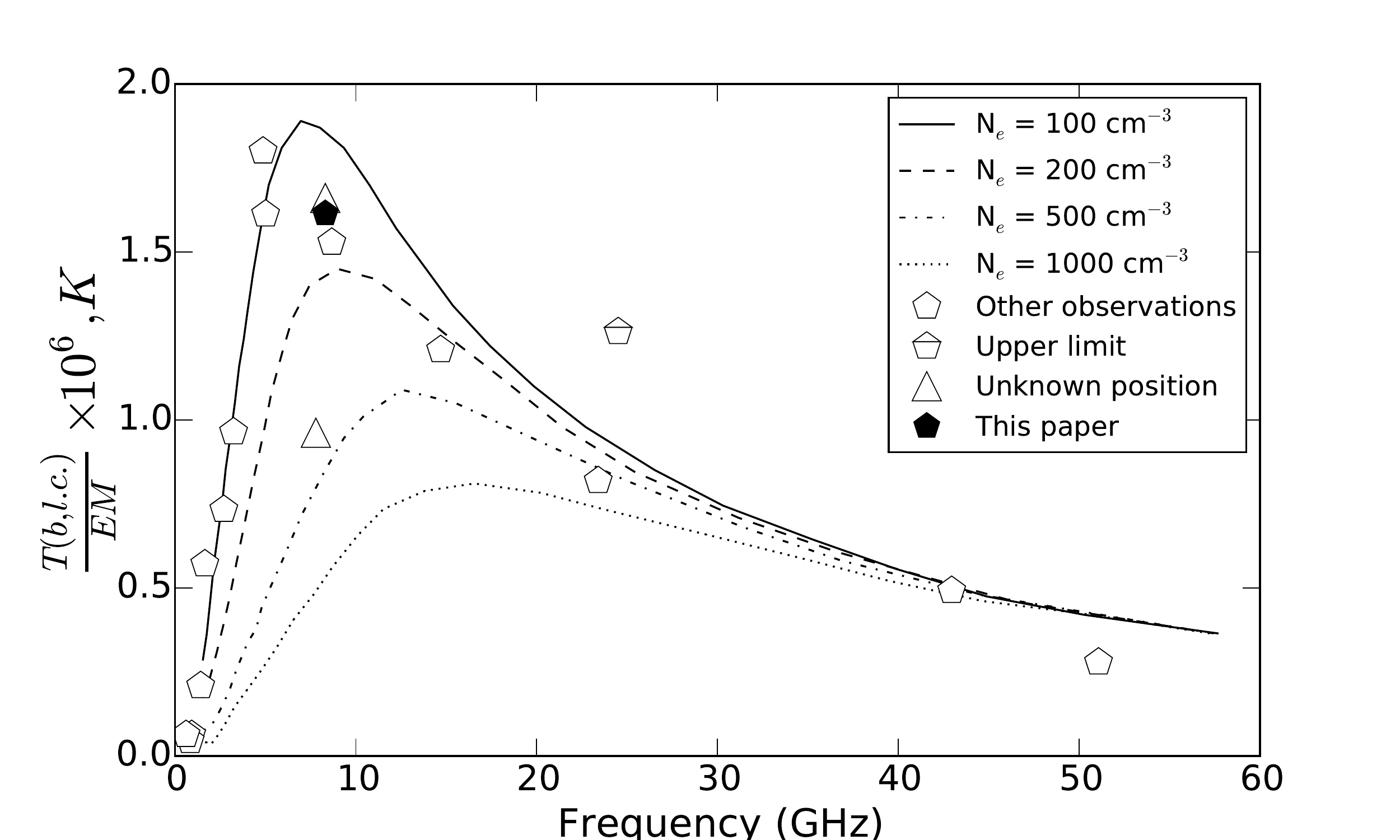}
\caption{
Brightness temperature at the center of a line as a function of frequency \citep{soro},
overlaid with peak antenna temperatures (Table~\ref{tab:lit_ref}) of Hydrogen RRLs towards Ori\,A observed at different frequencies.
}
\label{fig:OrionA_soro_lit}
\end{center}
\end{figure}

Fig.~\ref{fig:OrionA_soro_lit} is replotted from \cite{soro}, with observational data points added by taking the line brightnesses from Table~\ref{tab:lit_ref} divided by the emission measure of the {\sc Hii} region \citep[EM of $4.0\times 10^6$\,cm$^{-6}$\,pc taken from][]{1981MNRAS.196..889M}, and shows the variation of RRL brightness with electron density. All the lines denoted by pentagons are observed at similar positions, and the filled pentagon is the value taken from our map. Triangles denote observations with no position given. The model curves show that we should expect significant changes in the relative line brightnesses around 10\,GHz as the electron density increases from 
100~cm$^{-3}$ to 1000~cm$^{-3}$. Observations at these frequencies therefore present an opportunity to probe the electron density directly through RRLs, and large-scale mapping of RRLs at these frequencies may allow us to map electron density structure.

The data points in Fig.~\ref{fig:OrionA_soro_lit} generally lie near the model curves for electron densities of 100 and 200~cm$^{-3}$. 
Points lying significantly away from these tracks are at an uncertain position (7.8\,GHz) or an upper limit (24.5\,GHz). Estimates of the electron density towards Ori\,A generally lie at least an order of magnitude higher:
\cite{1984A&A...135..116S} calculated an electron density of (1$\pm$ 0.3)$\times 10^{4}$~cm$^{-3}$, based on Stark broadening of RRLs, and \cite{2007AJ....133..952G} suggest the electron density is generally around 2000~cm$^{-3}$ throughout Ori\,A, based on optical spectroscopy.

\citet{2007AJ....133..952G} found a layer of low-electron-density emission towards 
Ori\,A, lying in front of the main ionised region, and blueshifted. Fig.~\ref{fig:OrionA_soro_lit} shows that low-electron-density gas emits very strongly around 10\,GHz, so it is possible
that much of the H92$\alpha$ emission that we see arises in lower-density material.
\cite{1995ApJ...438..784W} modelled the dense layer of ionised material in Ori\,A as being 0.1--0.2~pc thick, so the denser gas might account for less than half of the 
$4\times 10^{6}$\,cm$^{-6}$pc emission measure; hence there could be a large amount of 
low-density gas to generate the RRLs.
To confirm this possibility,  a full spatial analysis of RRLs across frequencies and comparison to continuum maps would be required, falling outside the scope of this paper.

\begin{table*}
\caption{
Table containing Hydrogen RRL transition, rest frequency, peak brightness temperature (T$_{A}$), beam efficiency, line brightness temperature corrected for beam efficiency (T$_{L}$) and available publications with single dish observation towards Ori\,A in the H\textit{n}{$\alpha$} RRL.
The H109$\alpha$ T$_{L}$ appears to be have already been corrected for beam efficiency \citep{1978ApJ...226..869J}. The H64$\alpha$ T$_{L}$ had already been corrected for beam efficiency and with a smaller beam than our observations (42\arcsec), the data point in Fig~\ref{fig:OrionA_soro_lit} is an upper limit \citep{1997AA...327.1177W}.
Our 92$\alpha$ spectrum (R.A. 05:35:14.8, Dec. --05:22:32.4) is towards the overlap of the other H$(n)\alpha$ transitions.
}
\begin{center}
\begin{tabular}{ccccclcl}
\hline
Transition & Rest Frequency & T$_{A}$ & $\eta$ & T$_{L}$ & Reference\\
& (GHz) & (K) & & (K) \\
\hline
50 & 51.07 & 0.67 & 0.65 & 1.03& \cite{1977ApJ...214..699H} \\
53 & 42.95 & 0.98 & 0.65 $\pm$ 0.05& 1.97 & \cite{1988PASJ...40..581} \\
64 & 24.51 & --- & --- & 5.05 & \cite{1997AA...327.1177W}\\
65 & 25.40 & 1.54 & 0.47 & 3.28 & \citet{1970ApL.....5..157C} \\
76 & 14.69 & 2.95 & 0.70 & 4.21& \cite{1981MNRAS.196..889M} \\
91 & 8.67 & 4.28 & 0.70 & 6.11& \cite{2006ApJS..165..338Q} \\ 
92 & 8.31 & 4.65 & 0.70 $\pm$ 0.2 & 6.64 & \cite{Cesarsky} \\
92 & 8.31 & 4.57 & 0.70 & 6.53 & This paper \\
94 & 7.79 & 2.04 & 0.53& 3.85 & \cite{1967ApJ...149L..21G} \\ 
109 & 5.01 & --- & 0.65 & 6.45 & \cite{1978ApJ...226..869J} \\
110 & 4.87 & 5.40 & 0.75& 7.20 & \cite{1971ApJ...163..479D} \\
126 & 3.24 & 2.70 & 0.70 & 3.86 & \cite{1968AuJPh..21..149M} \\
134 & 2.70 & 2.20 & 0.75& 2.93 & \cite{1970AA.....4..244Z} \\
158 & 1.65 & 1.60 & 0.70 & 2.29& \cite{1969AuJPh..22..631M} \\
166 & 1.42 & 0.56 & 0.67 & 0.84 & \cite{1972MNRAS.159..129P} \\
192 & 0.92 & 0.17 & 0.67 & 0.25 & \cite{1972MNRAS.159..129P} \\
198 & 0.84 & 0.10 & 0.55 & 0.18 & \cite{1974ApJ...190...35Z} \\
220 & 0.61 & 0.17 & 0.67 & 0.25 & \cite{1972MNRAS.159..129P} \\
\hline
\end{tabular}
\end{center}
\label{tab:lit_ref}
\end{table*}

\subsection{Sagittarius A}

Sgr\,A has been mapped interferometrically in the H92$\alpha$ RRL using the VLA \citep{1993ApJ...418..235Z, 1993ApJS...86..133R, 2001AJ....121.2681L, 1997ApJ...481.1016L}, but the high resolution of these maps rules out simple comparisons to our data.

\begin{figure}
\begin{center}
\includegraphics[trim = 0cm 0cm 0cm 0cm ,clip=true, width=0.5\textwidth]{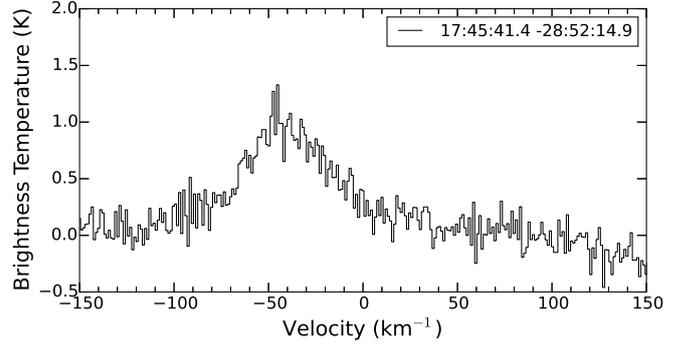}
\caption{Sample H92$\alpha$ spectra towards Sgr\,A.} 
\label{fig:SgrA_spectra}
\end{center}
\end{figure}

\cite{1976A&A....46..407P} and \cite{1980A&A....85...26P} used the 100-m Effelsberg telescope to map H85$\alpha$ and H109$\alpha$ RRL emission towards the Sgr\,A region. 
Continuum subtraction was applied on the Sgr\,A data set, although due to the strong Sgr\,A non-thermal emission affecting the entire band profile, there still remains some continuum emission seen in the maps as broadband emission in the southern peak ($\delta<-28^{\circ}55\arcmin$).
However, we see a `peanut' shape in our H92$\alpha$ map (Fig~\ref{fig:SGA}) similar to that seen by \citet{1976A&A....46..407P} in H85$\alpha$ (10\,GHz, 1.3$\arcmin$ beam);
our channel map (Fig~\ref{fig:SGA_channel}) and the spectrum towards the emission peak (Fig.~\ref{fig:SgrA_spectra}) are consistent with the finding of \citeauthor{1980A&A....85...26P} that the dominant H109$\alpha$ (5\,GHz, 2.6$\arcmin$ beam) emission occurs at $V_{LSR}\sim -40$\,km\,s$^{-1}$.

\subsection{On-the-fly mapping on the 70-m}

Our maps use a 30\arcsec\ row spacing, more finely-sampled than the Nyquist row spacing criteria suggested by \citet{mang}, resulting in oversampling of the rows.
Our scan rate implies an oversampling factor \citep[$n_{os}$ in Eq.\,10,][]{mang} close to 1.5 for Sgr\,A and 2 for Ori\,A.

Mapping a region the size of Ori\,A (18.9$\arcmin \times$17.5$\arcmin$ map) by position switching would take  about 25~hours, an order of magnitude more than the time taken to map it with OTF (Table~\ref{tab:regionInfo}).
The major contributor to the time requirement for position switching is the telescope motion overhead, as the telescope moves back and forth between the map position and the reference. 
From OTF mapping logs, each such movement takes 12--32\,s, 25\,s on average. 
Most of this time is required to accelerate, decelerate and settle the telescope, so it is likely not to be strongly dependent on the distance to the reference position.
Because mapping of the bright hydrogen RRLs only requires a few seconds' integration (compared to the tens of seconds of movement overhead), RRL mapping probably gains the most efficiency from OTF.

The ammonia inversion transitions in the 17--27\,GHz K-band are widely used to probe the temperature of cold dense gas; OTF may be used to map these lines.
A $10\arcmin\times 10\arcmin$ position-switching map with 60\,s integration on-source per position, mapped at half-beam spacing will take $\sim 29$~h and reach rms noise of $\sim 0.07$\,K. This sensitivity is likely to be useful for ammonia analysis. To map a similar area with OTF, spacing map rows by half a beam and sampling each row every third of a beam, using the minimum 3\,s integration time, takes 1.4~hours, achieving an estimated rms noise of 0.3~K.
Co-adding 12 repeated maps (taking into account that the OTF map is more densely sampled than the position-switched map), will yield an rms of $\sim 0.07$\,K in $\sim 17$~hours, about 0.6 of the time required for the position-switching map. 
Each row of the OTF map takes 150\,s, so the reference position is checked every 3.5\,min; 4\,min is the approximate maximum time between reference spectra at K-band.
K-band observations will require opacity and gain-curve corrections for the corrected temperature scale (T$_{A}^{*}$) which is applied after importing into \textsc{casa}. 
Smoothing using the Hanning function can be applied at a later stage.

For compact ammonia sources, observing efficiency can be increased by averaging the ends of the scan rows as an OFF spectrum in the data processing, removing the need for a reference scan.
This approach will require reformatting the raw file to allow \textsc{casa} to separate on-source spectra from OFF spectra for processing.

\section{Conclusions}
\label{sec:concl}
The 70-m Tidbinbilla radio telescope is now capable of on-the-fly (OTF) spectral line mapping. We have tested the new observing mode by mapping radio recombination line emission towards Orion~A and Sagittarius~A: Our data are consistent with published data towards these regions.

We have combined our measurement of the peak brightness towards Orion~A with other measurements from the literature over a wide range of frequency, and plotted the line brightnesses over the theoretical curves of \citet{soro}. The line brightness measurements are most consistent with electron densities of 
100--200\,cm$^{-3}$, much lower than usual estimates of the electron density of the Orion nebula. This discrepancy may reflect the more efficient RRL emission from lower-density gas. Comparison of RRL brightness over a wide frequency range has the potential to constrain the electron density of large ionised regions, but this requires large-scale mapping to allow different beams to be compared as well as to analyse spatial structure. OTF mapping, which is highly efficient at mapping RRLs, can enable this analysis.

Because of the long times required for the Tid-70m to switch between source and reference, large maps of bright lines (such as RRLs) are very efficiently mapped by OTF, while smaller maps of fainter lines (requiring longer integration times) do not gain as much. At K-band, the requirement to go to a reference position every $\sim 4$\,min sets a limit to the mapping efficiency that can be achieved. Even so, maps of ammonia inversion transitions should be significantly more efficient using OTF.


\section*{Acknowledgements}

DSS-43 is part of the Canberra Deep Space Communication Complex (CDSCC), which is managed by CSIRO Astronomy and Space Science.
G.W. thanks the director of CDSCC Dr. E. Kruzins for travel support and hospitality while part of this work was conducted, and Dr. Malte Marquarding for providing information related to \textsc{asap}.
CASA is developed at NRAO and under NRAO management with major contributions from ESO and NAOJ.
The original \textsc{auto\_spec} code was written by J.~E.~J. Lovell.
We thank the anonymous referee for the insightful comments, which have led to an improved paper. 








\appendix

\section{CASA process steps}
\label{app:casa}
The raw file from Tid-70m OTF is in \textsc{rpfits} format, imported by the \textsc{casa} task \textsc{sdsave}.
\textsc{sdlist} creates a list summary of the observations.
A plot of pointings before position correction can be generated by \textsc{sdplot} using parameter \textsc{pointing} in argument \textsc{plottype}.
An experimental script (\textsc{fixDirection\_v2.py}; see section \ref{sec:spec}), corrects the OTF pointings using the velocity taken from the \textsc{nmc} log files.
$a$ and $e$ are calculated for the new positions with the \textsc{asap} 
\textsc{recalc\_azel()} method on the scantable.
Opacity correction can be applied on the scantable through the \textsc{asap} \textsc{opacity} method.
\textsc{sdplot} is used again to confirm adjustments made to the pointings.
Calibration is completed by task \textsc{sdcal}, with the parameter \textsc{calmode} set to \textsc{quotient}.
Baseline subtraction can be applied through the command \textsc{sdbaseline}.
Spectra can be smoothed through the \textsc{sdsmooth} task, using the Hanning function by default.
To view the spectra, the task \textsc{sdplot} is used, setting parameter \textsc{plot type} to \textsc{grid}.
After calibration is completed, \textsc{sdsave} exports the data into a measurement set (MS) file.

\textsc{casa} task \textsc{sdimaging} generates the position-position-velocity cube, setting parameter \textsc{dochannel} to \textsc{true}.
The expandable parameters of \textsc{dochannel} are also required: \textsc{start}, \textsc{step} and \textsc{nchan}.
\textsc{start} gives the beginning of the cube in the velocity axis; \textsc{step} is the value for each channel; \textsc{nchan} is the number of channels in the velocity range.
Values for \textsc{start}, \textsc{step} and \textsc{nchan} are in units of \textsc{specunit}.
\textsc{Cell} is the size of each pixel: Values for the OTF maps are spacing between each scan (in Dec) and cell size along the scan derived by converting the scan rate (from the \textsc{nmc} log) from milli degree s$^{-1}$ to arc seconds s$^{-1}$, by the number of seconds per integration.
Image size is set by parameter \textsc{imsize}, based on the number of samples taken per scan and number of scans across source.
\textsc{sdimaging} by default averages both polarisations during the imaging process.

If baseline subtraction has not been applied, continuum subtraction can be used through task \textsc{imcont}.
A line free channel range must be specified, before running the command.
Integrated intensity images can be created by using the task \textsc{immoments}.
The image can be exported to a \textsc{fits} file by task \textsc{exportfits}.


\bsp	
\label{lastpage}
\end{document}